\documentclass{article}

 \PassOptionsToPackage{numbers, compress}{natbib}

\usepackage[final]{nips_2018}

\usepackage[utf8]{inputenc} 
\usepackage[T1]{fontenc}    
\usepackage{hyperref}       
\usepackage{url}            
\usepackage{booktabs}       
\usepackage{amsfonts}       
\usepackage{nicefrac}       
\usepackage{microtype}      
\usepackage{algorithm}
\usepackage{algorithmicx}
\usepackage{algpseudocode}
\usepackage{epsfig}
\usepackage{natbib}
\usepackage{graphicx}

\title{Robust Active Learning for Electrocardiographic Signal Classification}

\author{
 Xu Chen \\
 SAS Inc\\
  Cary, NC 27513 \\
\texttt{Xu.Chen@sas.com} \\
\And Saratendu Sethi\\
SAS Inc\\
Cary, NC 27513\\
\texttt{Saratendu.Sethi@sas.com} \\
}

\graphicspath{{C:/Action_writing/NIPSnew/}}
\begin{document}

\maketitle

\begin{abstract}
 The classification of electrocardiographic (ECG) signals is a challenging problem for healthcare industry. Traditional supervised learning methods require a large number of labeled data
which is usually expensive and difficult to obtain for ECG signals. Active learning is well-suited for ECG signal classification as it aims at selecting the best set of labeled data in order to maximize the classification performance.
Motivated by the fact that ECG data are usually heavily
unbalanced among different classes and the class labels are noisy as
they are manually labeled, this paper proposes a novel solution
based on robust active learning for addressing these challenges. The
key idea is to first apply the clustering of the data in a low
dimensional embedded space and then select the most information
instances within local clusters. By selecting the most informative
instances relying on local average minimal distances, the algorithm
tends to select the data for labelling in a more diversified way. Finally, the
robustness of the model is further enhanced by adding a novel noisy
label reduction scheme after the selection of the labeled data.
Experiments on the ECG signal
classification from the MIT-BIH arrhythmia database demonstrate the
effectiveness of the proposed algorithm.
\end{abstract}

\section{Introduction}
Many advanced machine learning algorithms require massive amount of
labeled data to demonstrate the full potential of the techniques. As
annotated data is usually expensive to obtain and requires the
expertise of people who are experts to label them, the goal of
active learning (AL)\cite{ICDM2016AL}\cite{NIPS2008AL}\cite{multiclassus}\cite{querybycommittee}\cite{expectedmodelchange} \cite{cvpr2013} \cite{icml2016} \cite{icml2017} is to
facilitate the data collection process by automatically selecting
the best data sample to label. Frequently, ECG signal classification often faces the problem of
unbalanced classes because naturally normal examples which
constitute the majority class in classification problems are massive
amount \cite{ECG}. For active learning, it is usually
important to predict and query for the minority classes, namely, abormal heart beats in ECG signals. Another
cause for class imbalance problem in ECG signal classification is the limitations due to cost,
difficulty or privacy on collecting instances of abnormal heart beats. The
work in \cite{CIKM07} proposed an efficient SVM based active
learning selection strategy to address the active learning for
unbalanced classes utilizing the information of support vectors for
querying the data in classification boundaries for binary
classification. However, \cite{CIKM07} did not utilize the important information of the clustering
structures of the unlabeled data for better selection of the data instances. More recently, the work in
\cite{LAL2017} reports that the more unbalanced the classes are, the
furtherfrom  the optimum made by uncertainty
sampling is. In addition to that, noisy labels obtained from the selection process and manual labelling pose
another challenge for active learning as the noisy labels could
propagate through the learning process if they are not well treated.

 In this paper, we propose an effective
solution to address these challenges in a unified framework called
robust active label spreading (RALS) thanks to two features. First,
 by selecting the most informative instances based on an
information theoretical measure and manifold embedding within the local clusters, the
proposed RALS algorithm is capable of selecting the samples from different classes in a more diversified manner, which results in better prediction performance. Second, a novel
noise reduction scheme is applied to further boost the prediction performance.
\begin{figure}[t]
  \caption{The detailed block diagram of the proposed RALS
algorithm. }
\label{blockdg}
  \centering
    \includegraphics[width=0.8\textwidth]{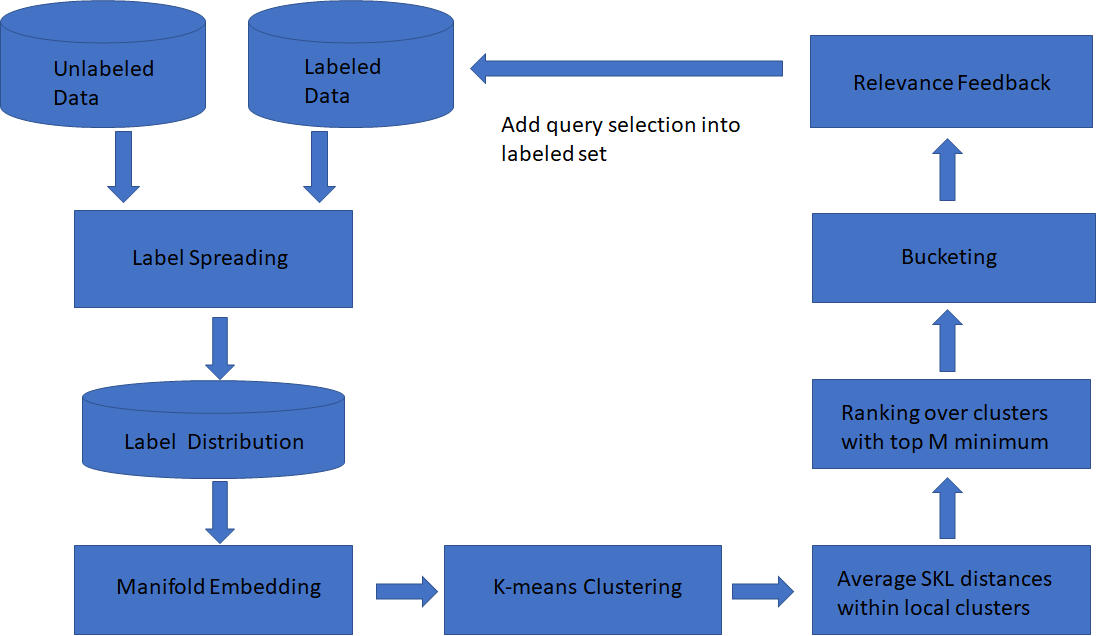}
\vspace{-1mm}
\end{figure}

\section{The Proposed Algorithm}

\paragraph{Dimensionality Reduction,  Clustering and Query Selection}
The first stage of RALS algorithm relies on label spreading \cite{NIPS2004}. The label spreading algorithm is a well known graph-based semi-supervised learning algorithm. It calculates the similarity measure and propagates the labels by the measure for prediction. It also
generates the label distribution matrix $ F^{N\times
C}$  which consists of the predicted probability for every class for each sample, where $C$ refers to the number of classes and $N$ refers to the number of
samples. In order to select the data from different classes, here t-Distributed
Stochastic Neighbor Embedding(t-SNE) \cite{tsne} is applied to the label
distribution matrix due to its good performance for high dimensional datasets.
As shown in Fig.1, once the dimensionality reduction is conducted, K-means clustering is applied to the label distribution matrix with reduced dimensionality ($N$ by $d$). The clustering and the query
selection  benefit from each other through the iterations, which
eventually boosts the classification performance. Given the dimensionality reduction and the clustering results, the
top $M$ $(M<<n)$ most informative points with their classified
labels are selected based on the
information theoretical measure as distance measure.
 Here we propose to utilize the local average minimum information theoretical distance within the clusters
for selection of the most informative data instances in a diversified manner.
Typically,
pair-wise symmetric kullback leibler
(SKL) divergences \cite{KL1951} within the local cluster
are calculated. The distances are then summed up for each sample and the top $M$ minimum distances are selected:\\
$\arg\min_{j}\frac{1}{N(d)}\sum_{i=1}^{N(d)}\sum_{k=1}^{c}(F_{kj}(t)\log\frac{F_{kj}(t)}{F_{ki}(t)}+F_{ki}(t)\log\frac{F_{ki}(t)}{F_{kj}(t)})$\\
where $F_{kj}$ represents the discrete probability distribution for
the $j$th sample and the $k$th class in the classification function.
$N(d)$ represents the the total number of instances in the $d$th
cluster. The above optimization
guarantees to select the the instances with the minimum distance
within the local clusters. Compared to the selection of data instances with minimum
distance globally, the proposed algorithm tends to diversify the top
selections over different clusters representing different classes.
Furthermore, we leverage the bucketing technique \cite{bucketcv}
which has been successfully applied to computer vision problems to ensure
that the samples from minority classes are selected. Specifically,
 the selected number of data instances Q needs to satisfy $[\frac{M}{C}]<Q<
 [\frac{M}{C}]+1$. Namely, if more than $ [\frac{M}{C}]+1$  samples within the same cluster are selected, the current cluster is skipped and the query goes to the next data in the ranking list, which results in better
classification performance.
\paragraph{Noisy Label Reduction}
Moreover,
 a novel noisy label reduction relying on an effective
confidence score measure is proposed based on the criteria of best vs second
best (BSVB) to  enhance the active
learning performance. Typically, for each selected data sample after ranking, the
ratio of the largest estimated class probability to the second largest
estimated class probability is calculated where this information can be retrieved from the label distribution matrix. Subsequently, the ratio is
compared to the user set threshold. The selected data are added into the labeled set  if the ratio is larger than the
threshold as:
$
\frac{P(\hat{y}|c_a)}{P(\hat{y}|c_b)}>\Delta,
$
where $\hat{y}$ represent the estimated label, $c_a$ and $c_b$  represent the class with the highest and the second highest
label distribution. $\Delta$ is the threshold which represents the
confidence level of the users.  Therefore, by adding the
estimated labels passed from the noise reduction step into the
labeled dataset, the noisy labels in the selection are significantly
reduced. The new augmented labeled
dataset after adding the selected data samples are applied to label spreading algorithm again to learn the
next enhanced model. The full algorithm is illustrated as follows.
\begin{algorithm}

 \begin{algorithmic}[1]

 \Procedure{\textbf{Algorithm I}: Robust Active Label Spreading}{$x_i$, $y_i$}
 \While{$m\leq \frac{\lambda-L}{M}$} \Comment{$\lambda$ is the total number of labeled data, $L$ is the total number of initial labeled data
 and $M$ is the number of selected labeled data in each iteration}
 \State Run label spreading algorithm \cite{NIPS2004},  output  label
 distribution matrix $F$.
 \State Apply t-SNE  on $F\rightarrow\tilde{F}$ for dimensionality reduction.  Conduct K-means clustering on $\tilde{F}$ where the
 number of classes is set to be the number of classes in the labeled dataset.
 \State Calculate the local minimum distance within each clusters relying on symmetric KL divergence $\arg\min_{j_{1},\ldots,j_{M}}\frac{1}{N(d)}\sum_{i=1}^{N(d)}\sum_{k=1}^{c}(F_{kj}(t)\log\frac{F_{kj}(t)}{F_{ki}(t)}+F_{ki}(t)\log\frac{F_{ki}(t)}{F_{kj}(t)})$
 \State Utilize the  bucketing technique to ensure the selected samples from the minority classes.
 \State Apply BVSB criteria for nosiy label reduction.  Add the selected samples $x_{j_{1}}, \ldots,x_{j_{M}}$ and estimated labels $\hat{y}_{j_{1}}, \ldots,\hat{y}_{j_{M}}$  to  labeled dataset and repeat steps (3)-(7).
 \State \textbf{return} $j_{1},\ldots,j_{M}$  \Comment{ $j_{1},\ldots,j_{M}$ are the selected label indices}.
 \EndWhile
 \EndProcedure
\end{algorithmic}
\end{algorithm}
\begin{figure}[t]
 \caption{The comparision of the precision recall curves with the uncertainty sampling (US)\cite{USKDD2009}, USDM \cite{CIKM07} and the propose method (RALS) on ECG dataset for the minority classes including Class A (Fig.2(a)) and LB (Fig.2(b)).The comparision of the total accuracy (Fig.2(c)) and the average accuracy (Fig.2(d)) vs the labeled samples for all the six classes of ECG data.}
\label{blockdg}
  \centering
    \includegraphics[width=0.8\textwidth]{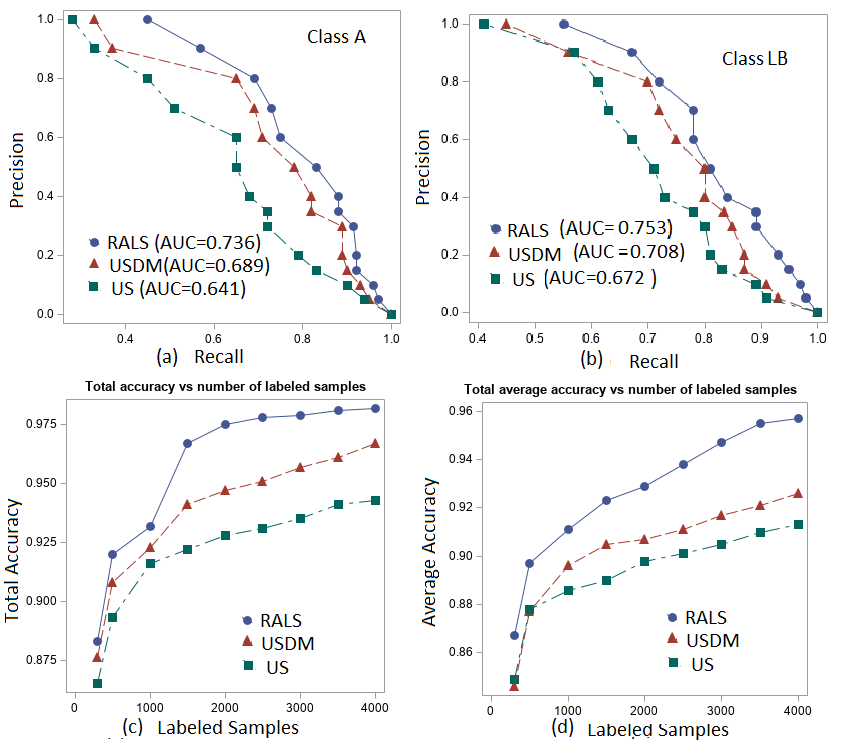}
\vspace{-1mm}
\end{figure}
\begin{table}
\begin{tabular}{|c|c|c|c|c|c|c|c|}
  \hline
 Labeled   & SVMAL\cite{CIKM07} & USDM \cite{multiclassus} &  \textbf{Ours}  & Labeled & SVMAL\cite{CIKM07} & USDM \cite{multiclassus} & \textbf{Ours} \\
 \hline
 N (25)  & 63\% & 67\% & 71\% & RB (25)  & 66\% & 70\% & 73\% \\
  \hline
 N (75) & 79\% & 83\% &  85\%  & RB (75)  & 71\% & 74\% &  78\% \\
  \hline
 N (125) & 82\% & 87\% & 89\% & RB (125)  & 74\% & 78\% & 82\% \\
  \hline
 N (175) & 85\%  & 87\% &  90\% & RB(175)  & 78\% & 82\% &  87\% \\
  \hline
  A (25)  & 68\% & 71\% &  73\% & LB (25) & 68\% & 72\% &  76\% \\
  \hline
 A (75) & 72\% & 73\%  & 76\%  & LB (75) & 72\% & 76\% & 78\% \\
  \hline
 A (125) & 76\% & 80\% & 82\% & LB (125) & 76\% & 83\% &  86\% \\
  \hline
 A (175) & 82\% & 83\% &  85\% & LB (175) & 78\%& 85\% & 89\%\\
  \hline
\end{tabular}
\caption{Comparison of the accuracy among  \cite{CIKM07},  USDM \cite{multiclassus} and the
proposed RALS algorithms for ECG dataset when using N, A, RB, LB as
the positive class for binary classification respectively.}
\vspace{-1mm}
\end{table}

\section{Experiments on Real ECG Data}
The proposed algorithm is evaluated on the real ECG data from the MIT-BIH arrhythmia database \cite{MITdatabase}.
The considered beats include the following six classes: normal sinus rhythm (N), atrial prematuring beat (A), ventricular premature beat (V),
right bundle branch block (RB), paced beat (P), and left bundle branch block (LB). The total 21344 beats were selected and preprocessed from the recordings of 20 patients as \cite{ECG}  including
12338 N, 344 A, 2194 V, 1982 RB, 3498 P and 988 LB beats.
Subsequently, a discrimative subset of features are extracted as described in \cite{ECG}: 1) ECG morphology features
and 2) three ECG temporal features.  The segmented ECG cycles are normalized to the same periodic legnth. The total feature dimension is 303 and is reduced to 6 dimensions using t-SNE. In label spreading algorithm, RBF kernel is chosen to
calculate the affinity matrix. The gamma parameter in the RBF kernel is
 0.25. The weight parameter
$\alpha$ is  0.2 .
100 data samples of the selected labeled beats is added in each iteration. The threshold $\Delta$ is 100.

The RALS algorithm is competed with two state-of-the-art active learning
methods: SVM based active learning (SVMAL) described in \cite{CIKM07} and  USDM described in \cite{multiclassus}. s \cite{CIKM07} can only be applied to binary classification, the performance for
binary classification is firstly compared. Typically,  the
four beat classes N,  A, RB and LB are selected as the positive class respectively and
the rest five classes are considered as the negative class.
We initialize all methods with 25 beats
and in each iteration 50 beats are added in the learning stage.
 Table 1 shows the proposed RALS algorithm
consistently outperforms \cite{CIKM07} and \cite{multiclassus}
under unbalanced classes with a large margin. Specifically, with a small number of labeled data (25 labeled beats), RALS achieves at least 5 \% and 3 \% better precision performance compared to uncertainty sampling with maximum entropy \cite{CIKM07} and USDM
\cite{multiclassus} respecitvley. The performance gain is mainly attributed to the fact that RALS employs a
better selection approach within clusters and noisy label reduction. Therefore, it enables the selection of
informative data instances in a diversified manner. Subsequenlty, we evaluate the multi-class classification performance by comparing our algorithm with uncertainty sampling (US) based on maximum entropy \cite{USKDD2009} and USDM \cite{multiclassus}. In Fig.2, we demonstrate the precision and recall curve comparison with the uncertainty sampling (US), USDM \cite{CIKM07} and the propose method (RALS) for ECG dataset for the minority classes including Class A and Class LB. The area under the curve (AUC) for RALS  is 0.736 for Class A and 0.753 for Class LB as the best method out of the three algorithms. Fig.2(c) and Fig2(d) plot the comparison of the total accuracy and the average total accuracy vs the labeled samples using the uncertainty sampling (US) \cite{USKDD2009},  USDM \cite{multiclassus} and the proposed RALS method. RALS is the top performer for both evaluation metrics again, which confirms the advantages of RALS in selecting diversified samples and reducing mislabeling risks.

\section{Conclusions}
The proposed RALS algorithm optimizes the selection process for labeled data
by selecting the data instances with
the minimum average distances in the local clusters and choosing
the candidates relying on the bucketing technique. The proposed
noise reduction scheme further reduces the noise labels and enhances
the classification performance. We applied RALS to the real ECG
signal classification and achieved superior performances over the
state-of-the-art approaches.\clearpage \small
\bibliographystyle{unsrt}
\bibliography{ml4h_final1}
\end{document}